\documentclass[preprint,showpacs,superscriptaddress,longbibliography]{revtex4-1}
\usepackage[dvipsnames]{xcolor}
\usepackage{diagbox}
\usepackage{amsmath}
\usepackage{amsfonts}
\usepackage{physics}
\usepackage[space]{grffile}
\usepackage{natbib}
\usepackage{braket}
\usepackage{float}
\usepackage{color}
\usepackage{blindtext}
\usepackage{comment}

\def\327{Sr$_3$Ir$_2$O$_7$}
\def\214{Sr$_2$IrO$_4$}
\def\TN{$T_\textrm{N}$}

\begin{document}

\title{Observation of excitons bound by antiferromagnetic correlations}

\author{Omar Mehio}
\thanks{These authors contributed equally to this work.}
\affiliation{Institute for Quantum Information and Matter, California Institute of Technology, Pasadena, CA 91125}
\affiliation{Department of Physics, California Institute of Technology, Pasadena, CA 91125}
\affiliation{Current address: Kavli Institute at Cornell for Nanoscale Science, Cornell University, Ithaca, NY 14853}

\author{Yuchen Han}
\thanks{These authors contributed equally to this work.}
\affiliation{Institute for Quantum Information and Matter, California Institute of Technology, Pasadena, CA 91125}
\affiliation{Department of Physics, California Institute of Technology, Pasadena, CA 91125}

\author{Xinwei Li}
\affiliation{Institute for Quantum Information and Matter, California Institute of Technology, Pasadena, CA 91125}
\affiliation{Department of Physics, California Institute of Technology, Pasadena, CA 91125}

\author{Honglie Ning}
\affiliation{Institute for Quantum Information and Matter, California Institute of Technology, Pasadena, CA 91125}
\affiliation{Department of Physics, California Institute of Technology, Pasadena, CA 91125}

\author{Zach Porter}
\affiliation{Linac Coherent Light Source and Stanford Institute for Materials and Energy Sciences, SLAC National Accelerator Laboratory, Menlo Park, California 94025, USA}
\affiliation{Materials Department, University of California, Santa Barbara, CA 93106}

\author{Stephen D. Wilson}
\affiliation{Materials Department, University of California, Santa Barbara, CA 93106}

\author{David Hsieh}
\email[Author to whom the correspondence should be addressed: ]{dhsieh@caltech.edu}
\affiliation{Institute for Quantum Information and Matter, California Institute of Technology, Pasadena, CA 91125}
\affiliation{Department of Physics, California Institute of Technology, Pasadena, CA 91125}

\maketitle
\noindent{\textbf{Abstract}}\\
\textbf{Two-dimensional Mott insulators host antiferromagnetic (AFM) correlations that are predicted to enhance the attractive interaction between empty (holons) and doubly occupied (doublons) sites, creating a novel pathway for exciton formation. However, experimental confirmation of this spin-mediated binding mechanism remains elusive. Leveraging the distinct magnetic critical properties of the Mott antiferromagnets \214 and \327, we show using time-resolved THz spectroscopy that excitons only exist at temperatures below where short-range AFM correlation develops. The excitons remain stable up to photodoping densities approaching the predicted excitonic Mott insulator-to-metal transition, revealing a unique robustness against screening. Our results establish the viability of spin-bound excitons and introduce opportunities for excitonic control through magnetic degrees of freedom.}


\newpage
\noindent{\textbf{Introduction}}\\
Excitons are bound states that form between photoexcited electrons and holes in insulating compounds. Conventionally, the bound state forms due to the attractive Coulomb interaction between the electron and hole. In two-dimensional (2D) Mott antiferromagnets, however, strong interactions between spin and charge degrees of freedom create novel pathways for bound state formation \cite{clarke_particle-hole_1993, huang_spin-mediated_2020, grusdt_pairing_2022, bohrdt_dichotomy_2022, lenarcic_charge_2014, lenarcic_ultrafast_2013, shinjo_density-matrix_2021, wrobel_excitons_2002, bohrdt_strong_2022}. When a doped carrier in the form of an empty (holon) or doubly occupied (doublon) site hops through the lattice, it disrupts the antiferromagnetic (AFM) motif, leaving in its wake a string of flipped spins that creates a confining potential (Fig. 1a). However, if a second charge carrier moves in tandem with the first, it partially restores the disruption, limiting spin flips to the length of string connecting the pair (Fig. 1b). The correlated motion of a holon and doublon represents the formation of a bound state known as a Hubbard exciton (HE). Unlike conventional semiconductor excitons that bind through Coulomb attraction, HE formation is driven by the competition between the kinetic energy of the charge carriers and the AFM exchange energy \cite{clarke_particle-hole_1993, huang_spin-mediated_2020, grusdt_pairing_2022, bohrdt_dichotomy_2022, bohrdt_strong_2022}.

Initially motivated as a mechanism for Cooper pair formation in high-temperature superconductors, this string-mediated pairing mechanism has been extensively studied from a theoretical point of view, revealing that HEs host rich internal structures \cite{clarke_particle-hole_1993, huang_spin-mediated_2020, grusdt_pairing_2022, bohrdt_dichotomy_2022, lenarcic_charge_2014, lenarcic_ultrafast_2013, shinjo_density-matrix_2021, wrobel_excitons_2002, bohrdt_strong_2022}, string-like excitation modes \cite{grusdt_pairing_2022}, and decay dynamics that depend sensitively on AFM correlations \cite{lenarcic_ultrafast_2013}. These theories have been experimentally borne out in atomic quantum simulators, in which spin polaron formation \cite{koepsell_imaging_2019, ji_coupling_2021}, string pattern formation \cite{chiu_string_2019}, and magnetically-mediated bound states \cite{hirthe_magnetically_2023, bohrdt_strong_2022} have been directly observed. In the solid state, however, the viability of spin-mediated pairing remains an open question. Spectral and dynamical signatures of HEs have been reported in several quasi-2D Mott antiferromagnets \cite{Mehio_Hubbard_2023, terashige_doublon-holon_2019, alpichshev_confinement-deconfinement_2015, novelli_ultrafast_2012, lovinger_influence_2020,gossling_mott-hubbard_2008}. However, because holons and doublons are oppositely charged, whether the exciton binding is primarily mediated by Coulomb attraction or AFM correlations remains an outstanding question. The importance of this question is highlighted by recent studies in one-dimensional copper oxide superconductors, in which the Coulomb interaction was found to be dominant in Cooper pair formation despite the like-charged nature of the charge carriers \cite{chen_anomalously_2021}.  

Disentangling these two contributions is challenging because spin-mediated binding should in principle occur even in the presence of short-range AFM correlations, which can persist well above the long-range magnetic ordering temperature in quasi-2D materials. Ideally, these effects can be distinguished in a material host that possesses strongly tunable short-range AFM correlations at temperatures irrelevant to thermal excitonic dissociation, yet retains its electronic properties throughout. The Ruddlesden-Popper series of square-lattice iridates Sr$_{n+1}$Ir$_{n}$O$_{3n+1}$ presents such an opportunity. The low energy electronic structure of both the $n = 1$ and $n = 2$ members consists of a completely filled band of spin-orbital entangled pseudospin $J_{eff} = 3/2$ states and a narrow half-filled band of $J_{eff} = 1/2$ states, which splits into lower and upper Hubbard bands (LHB and UHB) due to on-site Coulomb repulsion \cite{kim_novel_2008}. Accordingly, they possess very similar optical responses, featuring two prominent peaks marking the LHB $\rightarrow$ UHB transition ($\alpha$ peak) and the $J_{eff} = 3/2$ $\rightarrow$ UHB transition ($\beta$ peak) \cite{moon_temperature_2009, seo_infrared_2017}. Yet their spin subsystems belong to distinct universality classes, yielding stark differences in their AFM correlations near criticality. In single-layer \214, $J_{eff} = 1/2$ moments within the Ir-O planes are coupled through strong 2D Heisenberg interactions \cite{kim_magnetic_2012}. While a transition from long-range ordered AFM insulator to paramagnetic insulator occurs across a Néel temperature $T_{N} = 230$ K \cite{cao_weak_1998}, short-range AFM correlations persist to much higher temperatures - exceeding 100 lattice sites 20 K above \TN\ \cite{fujiyama_two-dimensional_2012} (Fig. 1c) - and the magnon spectrum survives even in the paramagnetic phase \cite{gretarsson_two-magnon_2016}. In contrast, \327 is composed of bilayers of Ir-O planes, creating additional out-of-plane exchange interactions \cite{kim_large_2012}. These interactions give rise to a collinear AFM ordered ground state \cite{kim_large_2012, moretti_sala_evidence_2015, mazzone_antiferromagnetic_2022, suwa_exciton_2021} with critical properties described by the 3D Ising universality class \cite{Vale_2019}. Above the AFM insulator to paramagnetic insulator transition ($T_{N}$ = 285 K), short-range AFM correlations rapidly disappear (Fig. 1d) along with the magnon spectrum \cite{kim_large_2012, gretarsson_two-magnon_2016}. Comparing excitonic responses in the critical region of both compounds can thus clarify the role of AFM correlations in HE formation.

Recently, a metastable population of HEs was observed in \214 using time-resolved time-domain THz spectroscopy (tr-TDTS) \cite{Mehio_Hubbard_2023}. In this experiment, a short pulse of pump photons resonant with the $\alpha$-peak excites a transient holon-doublon plasma characterized by a metallic Drude response. A short time later, these free carriers bind to form excitons, evidenced by a transfer of THz spectral weight from the Drude peak to a finite energy peak corresponding to an intra-excitonic transition. This clear spectral distinction between excitonic states and free carrier states makes tr-TDTS a powerful tool for studying excitonic dynamics \cite{Mehio_Hubbard_2023, kaindl_transient_2009, kaindl_ultrafast_2003}. On the other hand, studies of inter-excitonic transitions suffer from ultrafast bandgap renormalization effects following the photo-excitation, which severely obscure the distinction between the excitonic and free carrier response \cite{zhang_stability_2016, Mehio_Hubbard_2023}. Moreover, in these tr-TDTS studies, the excitons form after the photo-excited electrons relax towards the band edge through scattering processes with phonons, magnons, and other bosonic excitations, creating access to dark exciton states that could not otherwise be studied optically \cite{kaindl_ultrafast_2003, zhang_stability_2016}.

Here, we performed a comparative study of \214 and \327 using similar tr-TDTS measurements in reflection geometry on (001) oriented single crystals. We first demonstrate that HEs are metastable excitations of \327 by observing the transient formation of an intra-excitonic resonance. We then analyze the energetic properties of the intra-excitonic resonances in both \327 and \214 as a function of temperature. We find that while the excitons are stable at all temperatures in \214, the intra-excitonic resonance rapidly disappears above \TN\ in \327. By comparing the dielectric properties of the photo-excited state below \TN\ against that of the equilibrium state above \TN\, we find that the HEs can exist in dielectric environments far more metallic than the conductivity enhancement caused by the thermal melting of AFM order, ruling out an electronic origin to the temperature dependence of the intra-excitonic resonance. Instead, we conclude that the HEs can only exist at temperatures below where short-range AFM correlation develops. Owing to the differences in their magnetic universality classes - 3D Ising for \327 and 2D Heisenberg for \214 - this temperature scale is closely tied to \TN\ for \327 but is laregely unaffected by \TN\ for \214. Finally, we find that the excitons remain unaffected by large photo-carrier densities (4 mJ/cm$^{2}$) approaching the theoretical excitonic Mott transition density, demonstrating a unique property of these magnetically-bound HEs. \\

\noindent\textbf{Experimental Methods}\\
\noindent\textit{Sample Growth}\\
Single crystals of Sr$_{3}$Ir$_2$O$_{7}$ were synthesized via a flux growth technique. High purity powders of SrCO$_3$, IrO$_2$, and La$_2$O$_3$ (Alfa Aesar) were dried, and stoichiometric amounts were measured out, employing a 15:1 molar ratio between SrCl$_2$ flux and target composition. Powders were loaded into a platinum crucible and reacted. Deionized water was used to dissolve the SrCl$_2$ flux and expose the crystals grown at the base of the crucible. Typical crystals had a plate-like habit with dimensions typically of 1 mm $\times$ 1 mm $\times$ 0.2 mm \cite{hogan_PhysRevLett.114.257203}.

Single crystals of Sr$_{2}$IrO$_{4}$ were grown using a flux technique from stoichiometric quantities of IrO$_{2}$, SrCO$_{3}$, which were mixed with the flux in a 5:1 molar ratio of SrCl$_2$:Sr$_2$IrO$_4$. The ground mixtures of powders were melted at 1370 \textdegree C in partially capped platinum crucibles. The soaking phase of the synthesis lasted for 5 h and was followed by a slow cooling at 6 \textdegree C/h to reach 850 \textdegree C. From this point, the crucible is brought to room temperature through rapid cooling at a rate of 100 \textdegree C/h. The (001) face of the crystals were polished to a mirror finish using diamond lapping paper with a grit size of 1 $\mu$m. Typical crystal size was 1 mm $\times$ 1 mm $\times$ 0.2 mm. \\

\noindent\textit{Time-Domain THz Spectroscopy}\\
In our tr-TDTS experiments, holon-doublon pairs are excited using a 100 fs (FWHM) pump pulse tuned above the LHB to UHB transition. After a variable time delay $t$, differential changes to the low-energy charge response are probed with a phase-locked broadband THz pulse (0.8 THz $\rightarrow$ 5.5 THz), whose electric field profile is measured in the time domain via electro-optic sampling (EOS) as a function of the recording time $t_{EOS}$. These traces are then converted to the frequency domain using a fast Fourier transform with respect to $t_{EOS}$.

The tr-TDTS setup was seeded by 800 nm, 35 fs pulses produced by a Ti:sapphire amplifier operating at 1 kHz, which was split into three arms. The first arm (3.5 mJ pulse energy) was sent into an optical parametric amplifier (OPA) that served as the tunable near-infrared (NIR) pump source. The OPA is tuned to 1250 nm (1 eV) in Sr$_3$Ir$_2$O$_7$ and 2050 nm (0.6 eV) in Sr$_2$IrO$_4$, both of which are well above the Mott gap energy in each material. The second arm (0.6 mJ pulse energy) was used to generate the broadband THz-frequency probe through optical rectification of the 800 nm pulses. The third arm (1 $\mu$J pulse energy) was reserved for the electro-optic sampling (EOS) gate pulse used to measure the THz electric field $E(t_{\textrm{EOS}})$. The temporal delay $t_{\textrm{EOS}}$ between the 800 nm EOS gate pulse and the THz probe was controlled with a motorized delay stage.

The TDTS setup was utilized in reflection geometry to measure the photo-induced changes to the optical spectra of Sr$_{2}$IrO$_{4}$ and \327. The THz probe beam was $s$-polarized and incident onto the (001) sample surface at a 30$^{\circ}$ angle of incidence. The pump beam was normally incident onto the sample and was cross-polarized with the THz probe. The transient pump-induced change to the THz electric field is defined as $\Delta E\left(t_{\mathrm{EOS}}, t \right) = E_{Pumped}\left(t_{\mathrm{EOS}}, t \right)-E_{Static}\left(t_{\mathrm{EOS}}\right)$, where $t$ is the relative delay between the pump pulse and the EOS gate pulse. To measure the pump-induced waveform $\Delta E\left(t_{\mathrm{EOS}}, t \right)$, $t$ is kept fixed and both the EOS gate pulse and the pump pulse are swept along the EOS time axis $t_{\mathrm{EOS}}$ using two motorized delay stages that are synchronized with each other. 

The differential measurement of the pump-induced waveform $\Delta E\left(t_{\mathrm{EOS}}, t \right)$ is enabled by mechanically chopping the NIR pump pulse at half the repetition rate of the probe pulse. The mechanical chopper served as the reference trigger for a lock-in amplifier, which filtered the EOS signal from a balanced photodiode. Such a detection technique yields a “pump-on” minus “pump-off” detection of the THz electric field. This technique can be operated with either one or two choppers and lock-in amplifiers. In the latter case, the static electric field can be detected simultaneously with the transient changes, ensuring that any spectral artifacts due to long-term drift are eliminated. The cost is a reduction in the signal-to-noise ratio by half. We utilized both methods in our measurements depending on the strength of the signal; both yielded identical results. When the single-chopper method was used, the static THz electric field and its transient changes were measured sequentially at each $t$ to ensure that there were no spectral artifacts. 

Two different configurations of detection and generation crystals were used in these measurements. The first utilized a 0.2 mm thick $<$110$>$ GaP crystal for generating the THz pulse and a 0.2 mm thick $<$110$>$ GaP crystal mounted on 1 mm thick $<$100$>$ GaP for EOS detection, yielding a bandwidth of 0.5 THz to 6 THz. The second utilized 1.0 mm thick $<$110$>$ ZnTe crystals for both generation and EOS detection, yielding a bandwidth of 0.3 THz to 2 THz. The entire tr-TDTS apparatus was enclosed in a N$_{2}$ gas-purged environment.\\

\noindent\textbf{Intra-excitonic transition in Sr$_{3}$Ir$_{2}$O$_{7}$}\\
We first investigate whether \327 also hosts a metastable HE population. A typical dataset from \327 is shown in Figure 2a, which was acquired with a pump photon energy of 1.0 eV and a pump fluence of 4 mJ/cm$^2$ in the long-range AFM ordered phase. A rise in the reflected THz field amplitude occurs immediately upon arrival of the pump pulse, followed by an exponential decay. During this time, several peaks emerge in the frequency domain. The narrow peaks above 2 THz can all be assigned to the set of lowest-energy infrared-active phonon modes reported in this compound (Fig. S1 \cite{sm}). On the other hand, the broad peak centered around 1 THz lies below both the known lowest-energy optical phonon modes \cite{ahn.moon_srep16} and the zone-center magnon gap ($\sim$90 meV) \cite{kim_large_2012}, ruling out a structural or magnetic origin. 

Several observations support an excitonic origin of this peak. First, this peak is absent in equilibrium (Fig. S2 \cite{sm}) and only appears after photo-excitation, as expected for an intra-excitonic transition \cite{kaindl_ultrafast_2003, kaindl_transient_2009}. Second, the full dielectric response of the differential THz electric field spectrum can be extracted through standard electrodynamic relations using the thin film approximation (Methods \cite{sm}). Figure 2b shows the pump-induced change to the real $\sigma_{1}(\omega)$ and imaginary $\sigma_{2}(\omega)$ parts of the optical conductivity at different $t$ in the vicinity of the 1 THz peak (white box in Fig. 2a). At $t$ = 0 there is an increase in both $\sigma_{1}(\omega)$ and $\sigma_{2}(\omega)$, consistent with the Drude response of a conducting holon-doublon plasma \cite{okamoto_photoinduced_2011, zhang_stability_2016, steinleitner_direct_2017}. But after about 1.6 ps, the response becomes predominantly Lorentzian, evidenced by $\Delta \sigma_{1}(\omega)$ exhibiting a peak centered around 1 THz and $\Delta \sigma_{2}(\omega)$ exhibiting a dispersive lineshape with a zero-crossing at the same frequency. This is consistent with a dipole-active transition between excitonic energy levels \cite{kaindl_ultrafast_2003, zhang_stability_2016, poellmann_resonant_2015} reminiscent of the 1.5 THz intra-excitonic transition found in \214 \cite{Mehio_Hubbard_2023}. Third, to quantitatively study the Drude to Lorentzian crossover, $\Delta \sigma_{1}(\omega)$ and $\Delta \sigma_{2}(\omega)$ were simultaneously fit to a sum of Drude and Lorentz oscillator functions at each $t$ (Methods, Fig. S3 \cite{sm}). Their respective spectral weights (SW) were extracted from the area under the fits to $\Delta \sigma_{1}(\omega)$ (Methods \cite{sm}). As shown in Figure 2c, the rise in Lorentzian SW coincides with the decay of Drude SW, signalling a population transfer from free to bound carriers, a hallmark of transient exciton formation \cite{Mehio_Hubbard_2023,kaindl_ultrafast_2003, kaindl_transient_2009, zhang_stability_2016, steinleitner_direct_2017}. Fourth, the zone-boundary magnon energy - a measure of local spin exchange couplings - is roughly 160 meV and 200 meV in \327 \cite{kim_large_2012} and \214 \cite{kim_magnetic_2012} respectively. This ratio of energies is comparable to the ratio between the intra-excitonic peak frequencies in \327 (1 THz) and \214 (1.5 THz \cite{Mehio_Hubbard_2023}), which is expected in a spin-mediated exciton binding scenario \cite{Mehio_Hubbard_2023, lenarcic_exciton_2015, lenarcic_ultrafast_2013, lenarcic_charge_2014, grusdt_pairing_2022}. Note that while the magnon spectrum of \214 is uniquely described by a weakly anisotropic Heisenberg model \cite{kim_magnetic_2012}, there are competing descriptions of the magnetic excitation spectrum of \327 ranging from highly anisotropic Heisenberg \cite{kim_large_2012}, to quantum dimer \cite{moretti_sala_evidence_2015}, to AFM excitonic insulator models \cite{mazzone_antiferromagnetic_2022}. Since the fitted values of the spin exchange differ substantially between these models, we chose the zone boundary magnon energy as a model-independent measure of spin exchange. \\

\noindent\textbf{Correlation-driven exciton formation}\\
To determine how these excitonic features evolve across \TN, we extracted the temperature dependence of the full dielectric response of both \327 and \214 in the vicinity of their intra-excitonic transition peaks. In \214, hallmarks of a transient Lorentzian component in both the $\Delta \sigma_{1}(\omega)$ and $\Delta \sigma_{2}(\omega)$ persist up to our highest measured temperature of 300 K ($T$/\TN = 1.3) (Figs 3a,b). In contrast, the Lorentzian component observed in \327 (Fig. 2) is very sensitive to temperature and becomes undetectable above \TN\ (Figs 3c,d). In the paramagnetic phase of \327, only a positive change in both real and imaginary parts of the optical conductivity is resolved at all time delays (Fig. S4). Although a Drude model does not perfectly fit these spectra (Methods \cite{sm}), the positive response indicates that the transient state is metallic. 

To obtain a detailed temperature dependence of the excitonic response in the critical region around \TN, we measured the maximum pump-induced change of the THz field ($\Delta E_{max}$) at a fixed $t_{EOS}$, where the static THz pulse reaches a maximum electric field. This captures the frequency-integrated photo-induced change in the THz electric field amplitude \cite{poellmann_resonant_2015}, thereby tracking the Lorentzian SW (Methods, Fig. S5 \cite{sm}). As shown in Figure 3e, for \214 the change in amplitude is relatively temperature-independent across \TN. But for \327, it undergoes a sharp superlinear increase upon cooling near \TN\ and then reaches a plateau, which is reminiscent of the temperature dependence of the AFM correlation length (Fig. 1d). These data strongly suggest that the viability of HEs depends crucially on the presence of short-range AFM correlations.

Although the Mott gap in both \214 and \327 stays intact in the paramagnetic phase \cite{moon_temperature_2009, king_spectroscopic_2013, song_magnetically_2018, foulquier_evolution_2023, cao_weak_1998}, there is a transfer of spectral weight from the Hubbard bands to in-gap states upon heating through \TN\ \cite{moon_temperature_2009,king_spectroscopic_2013, song_magnetically_2018, foulquier_evolution_2023} owing to their moderately correlated nature \cite{watanabe_theoretical_2014, hsieh_observation_2012}. Since this effect is more pronounced in \327, a natural question is whether the disappearance of the excitonic peak in \327 can be attributed simply to an enhanced conductivity above \TN. This scenario can be ruled out by analyzing the dielectric properties of \327 in the photo-excited state. Figure 2c shows that from roughly $t = 1$ ps to beyond 6 ps, there is a coexistence of Drude and Lorentz components, demonstrating that HEs are robust against a finite population of free carriers. At $t = 2$ ps where the Lorentzian SW is maximum, the instantaneous DC conductivity - extrapolated from the Drude component of the fit to $\Delta\sigma_{1}(\omega)$ - exceeds 220 $\Omega^{-1}$cm$^{-1}$ (Fig. 4a). This value is four times higher than the reported equilibrium DC conductivity at 300 K ($<$50 $\Omega^{-1}$cm$^{-1}$ \cite{cao_PhysRevB.66.214412}), suggesting that an increased conductivity above \TN\ alone cannot explain the disappearance of the excitonic response.\\

\noindent\textbf{Absence of an excitonic Mott transition}\\
To further examine the impact of dielectric screening on the excitonic peak, we performed transient THz conductivity measurements on \327 as a function of pump fluence. As shown in Figures 4b and 4c, there is no measurable change in the exciton peak over a four-fold increase in fluence up to 4 mJ/cm$^{2}$. The same behavior is observed in \214 (Fig. S6 \cite{sm}). This is in stark contrast to conventional Coulomb-bound exciton peaks, which exhibit significant red-shifting and linewidth broadening with increasing fluence due to screening \cite{chernikov_population_2015, zhang_stability_2016}. Above a critical photo-carrier density $d_{c}$ - approximately $1/a_{0}^{2}$, where $a_{0}$ is the excitonic Bohr radius \cite{klingshirn_semiconductor_2007} - where the binding energy vanishes, Coulomb-bound excitons undergo a Mott transition and dissociate into a conducting electron-hole plasma \cite{kaindl_transient_2009, klingshirn_semiconductor_2007, chernikov_population_2015}. In conventional 2D semiconducting systems, values for $d_c$ range from 1 $\times$ $10^{11}$ cm$^{-2}$ for GaAs quantum wells \cite{zhang_stability_2016} up to 1 $\times$ $10^{14}$ cm$^{-2}$ for WS$_{2}$ \cite{chernikov_population_2015}. In the case of \214 and \327, assuming the smallest possible excitonic Bohr radius where the holon and doublon are on neighboring Ir sites -  roughly 0.4 nm \cite{fujiyama_two-dimensional_2012} - we should expect that $d_c \sim$ 6 $\times$ $10^{14}$ cm$^{-2}$ in a Coulomb-bound scenario. However, up to our maximum fluence, which corresponds to a photo-carrier density ($\sim$1 $\times$ $10^{14}$ cm$^{-2}$; Methods \cite{sm}) close to this maximal $d_c$, we observe no signs of approaching a Mott transition - neither red-shifting nor broadening of the peak. This observation suggests that the excitons in \214 and \327 are unusually robust against the screening effects that drive the excitonic Mott transition in conventional semiconductors, further ruling out Coulomb attraction as the primary binding mechanism. \\


\noindent\textbf{Conclusions}\\
Our results collectively support the existence of spin-bound excitons in 2D Mott antiferromagnets that are stabilized by short-range AFM correlations. The predominance of spin-mediated over Coulomb-mediated carrier interactions in the Ruddlesden-Popper iridates may also be conducive to holon-holon or doublon-doublon binding, paving a possible path to Cooper pairing and high-temperature superconductivity \cite{grusdt_pairing_2022, wang_twisted_2011, kim_observation_2016}. The spin-mediated binding mechanism further introduces an opportunity to engineer excitonic properties via the magnetic properties of a material. The plethora of tools available for manipulating quantum magnets (e.g. tailoring short-range AFM correlations through lattice geometry \cite{li_magnetic_2021}, inducing phase transitions through applied magnetic fields \cite{seyler_spin-orbit-enhanced_2020, porras_pseudospin-lattice_2019}, or dynamically tuning spin exchange interactions through Floquet engineering \cite{mentink_ultrafast_2015, Chaudhary2019}) can all potentially be brought to bear on developing novel exciton-based technologies. \\

\noindent\textbf{Acknowledgments}\\
The authors thank Zala Lenarčič, Michael Buchhold, Eugene Demler and Victor Galitski for useful discussions. We thank S. J. Moon for sharing his equilibrium optical conductivity results. Terahertz spectroscopy measurements were supported by NSF Award DMR-2104833. D.H. acknowledges support for instrumentation from the Institute for Quantum Information and Matter (IQIM), an NSF Physics Frontiers Center (PHY-2317110). S.D.W. acknowledges partial support via NSF award DMR-1729489. This work used facilities supported via the U.C. Santa Barbara NSF Quantum Foundry funded via the Q-AMASE-i program under award DMR-1906325. 


%

\newpage
\begin{figure}[t]
\includegraphics[width=\textwidth]{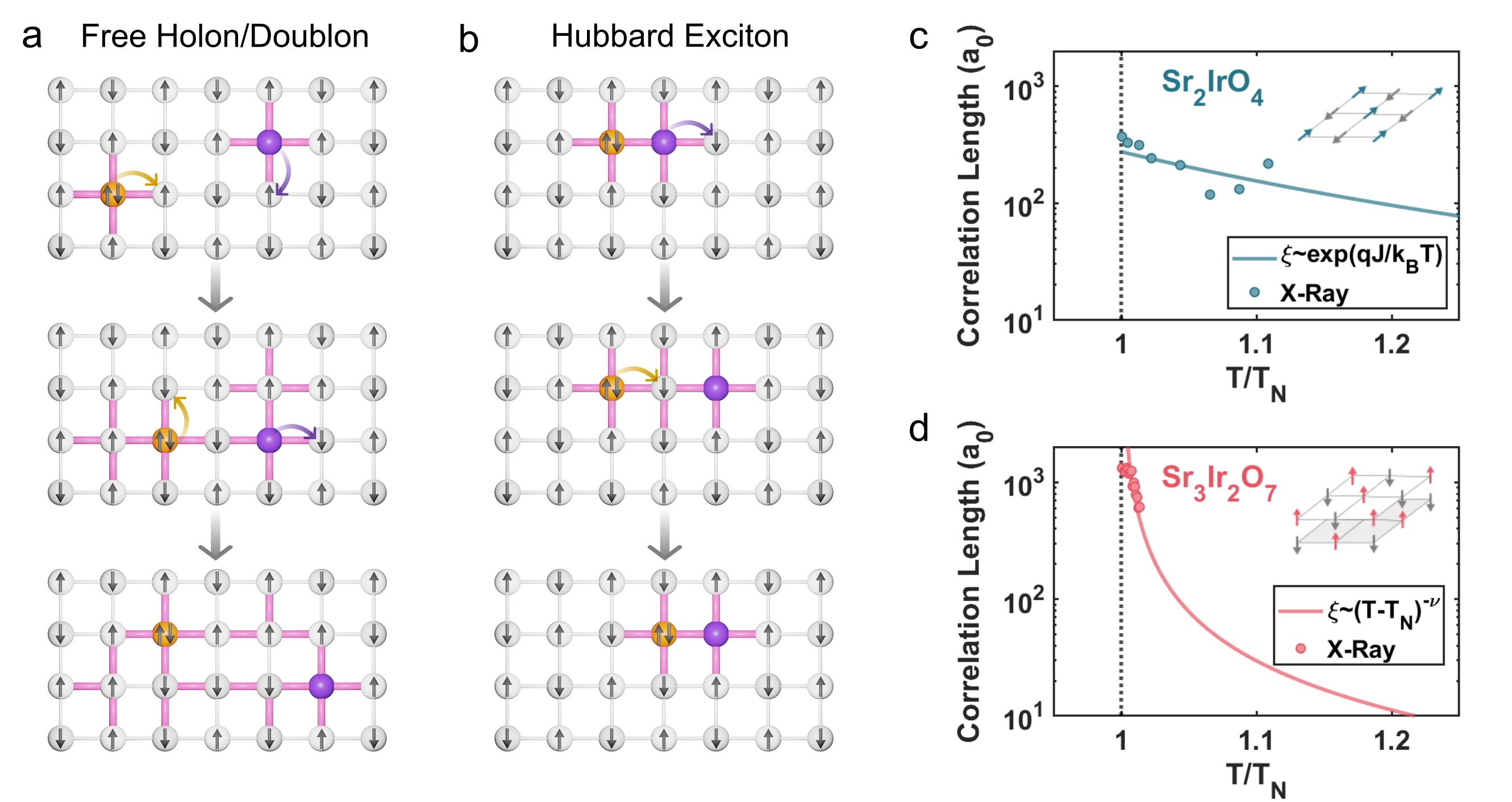}
\label{Fig1}
\end{figure}
\noindent
\textbf{Figure 1 $|$ Hubbard excitons and the magnetic critical properties of insulating Ruddlesden-Popper iridates.} 
\textbf{a} Schematic of a free holon (purple) and doublon (orange) moving through an AFM lattice. Top to bottom panels show spin configurations after successive hops (purple and orange arrows). The line of excited bonds (magenta) is the string defect. \textbf{b} Schematic of a HE moving through an AFM lattice. When the holon hops away from the doublon (top to middle panel), the length of the string defect increases. However, if the doublon coherently follows the holon (bottom panel), the string defect recovers its original length. \textbf{c,d} Measured temperature dependence of the spin correlation length of \214 \cite{fujiyama_two-dimensional_2012} (c) and \327 \cite{Vale_2019} (d) shown as points, where $a_0$ is the lattice constant. Solid lines are fits to the scaling law based on 2D Heisenberg (c) and 3D Ising (d) universality classes, with fit parameters provided in \cite{fujiyama_two-dimensional_2012,Vale_2019}. Insets show the long-range ordered spin configuration.

\newpage

\begin{center}
\begin{figure}
\includegraphics[width=\textwidth]{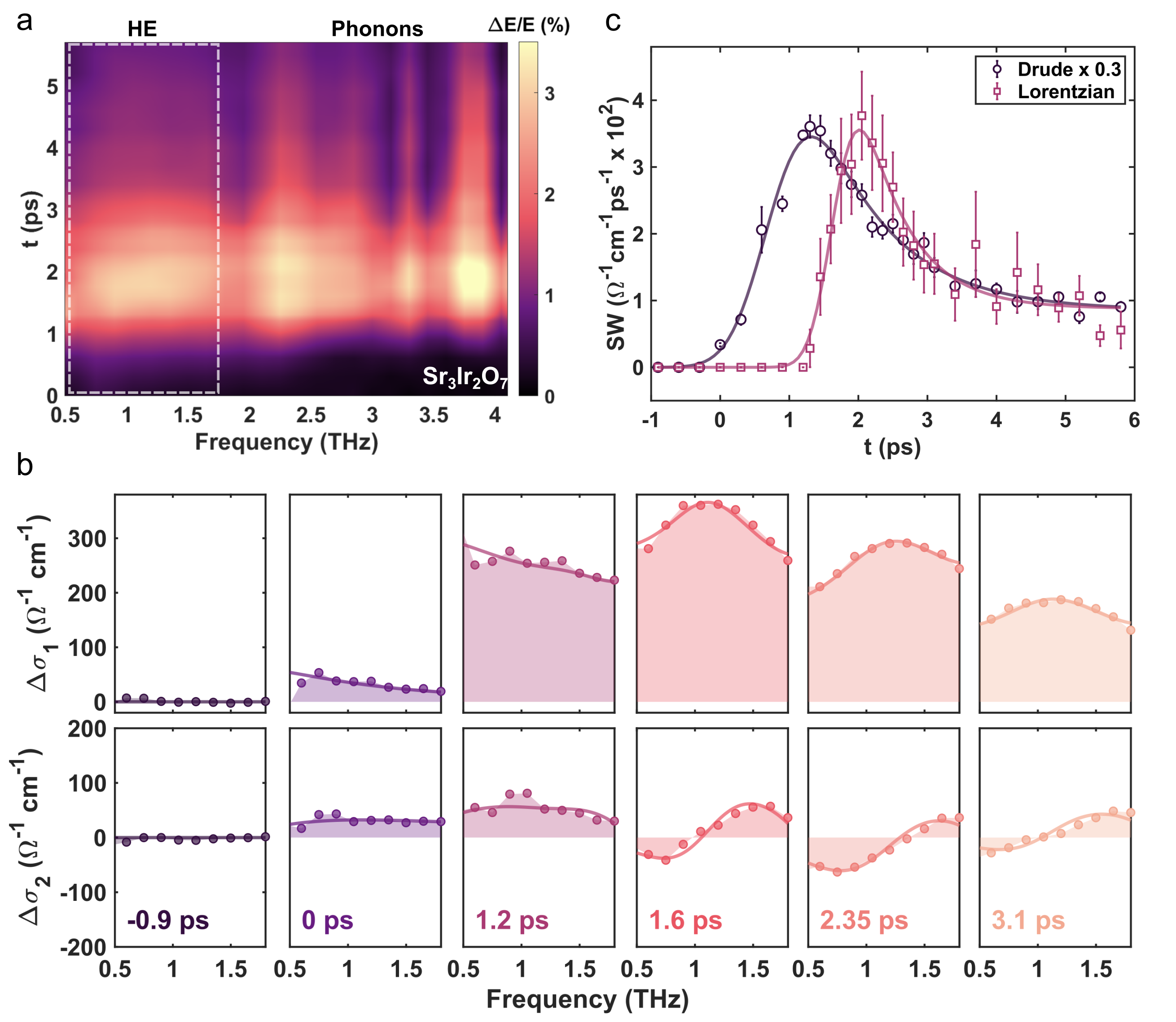}
\label{Fig2}
\end{figure}
\end{center}
\noindent
\textbf{Figure 2 $|$ Hubbard exciton dynamics in \327 at $T < $ \TN. }
\textbf{a} Pump-induced change of the THz spectrum as a function of pump-probe delay $t$. Colors represent the magnitude of $\Delta E/E$. The data were acquired at 80 K with a pump fluence of 4 mJ/cm$^{2}$. The dashed box indicates the frequency range used for the Drude-Lorentz fit shown in panels b-c. \textbf{b} $\Delta\sigma_{1}(\omega)$ (top panels) and $\Delta\sigma_{2}(\omega)$ (bottom panels) at various $t$, displayed as dots and shading, extracted from the differential THz spectra shown in panel a (Methods \cite{sm}). Fits to the Drude-Lorentz model (Methods \cite{sm}) are displayed as solid lines. \textbf{c} Fitted Drude and Lorentzian spectral weight versus $t$. Solid lines are fits to an exponential function (Methods \cite{sm}). Error bars are obtained from the standard deviation of the least-squares-fitting algorithm.

\newpage 

\begin{figure}
\includegraphics[width=0.62\textwidth]{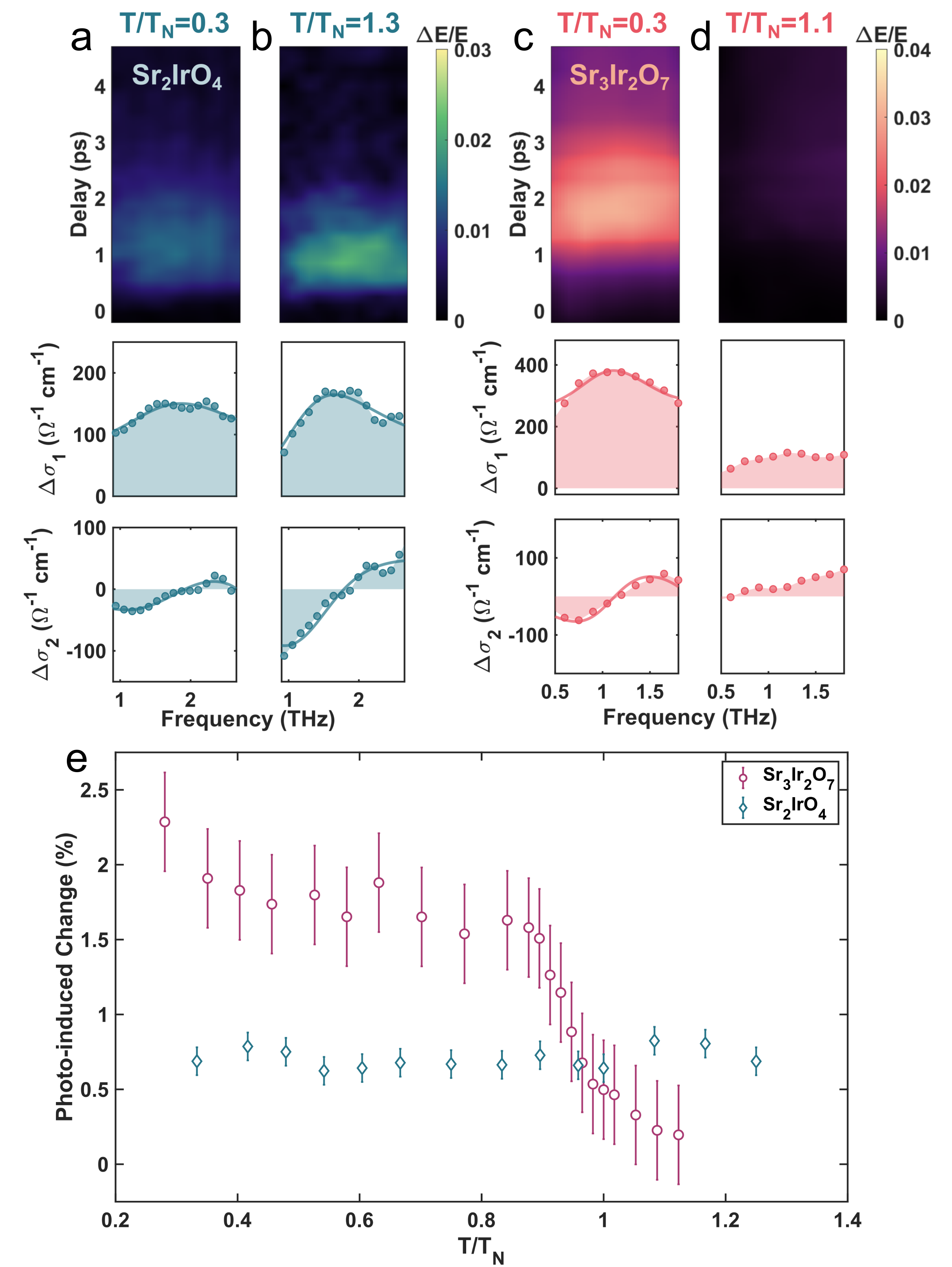}
\label{Fig3}
\end{figure}
\noindent
\textbf{Figure 3 $|$ Temperature dependence of the excitonic response in \214 and \327.} 
\textbf{a,b} $|\Delta E/E(\omega)|$ (upper panels), $\Delta\sigma_{1}(\omega)$ (middle panels) and $\Delta\sigma_{2}(\omega)$ (lower panels) of \214 at temperatures below (a) and above (b) \TN. The transient conductivity data shown for a representative delay of $t$ = 0.9 ps and are displayed as dots and shadings. Solid lines are least-squared fits to a Drude-Lorentz model (Methods \cite{sm}). The pump photon energy and fluence were 0.6 eV and 2 mJ/cm$^2$ respectively. \textbf{c,d} Same as panels a and b but for \327. In this case, the pump photon energy and fluence were 1 eV and 4 mJ/cm$^2$ respectively, and the transient conductivity data are shown at a representative delay of $t$ = 1.75 ps. \textbf{e} Temperature dependence of the frequency-integrated peak amplitude of $\Delta E/E(t_{EOS})$ for \214 (teal) and \327 (magenta) (Methods, Fig. S5 \cite{sm}). Error bars are the standard deviation obtained from 3 (for \327) or 4 (for \214) independent measurements.
\newpage

\begin{figure}
\includegraphics[width=\textwidth]{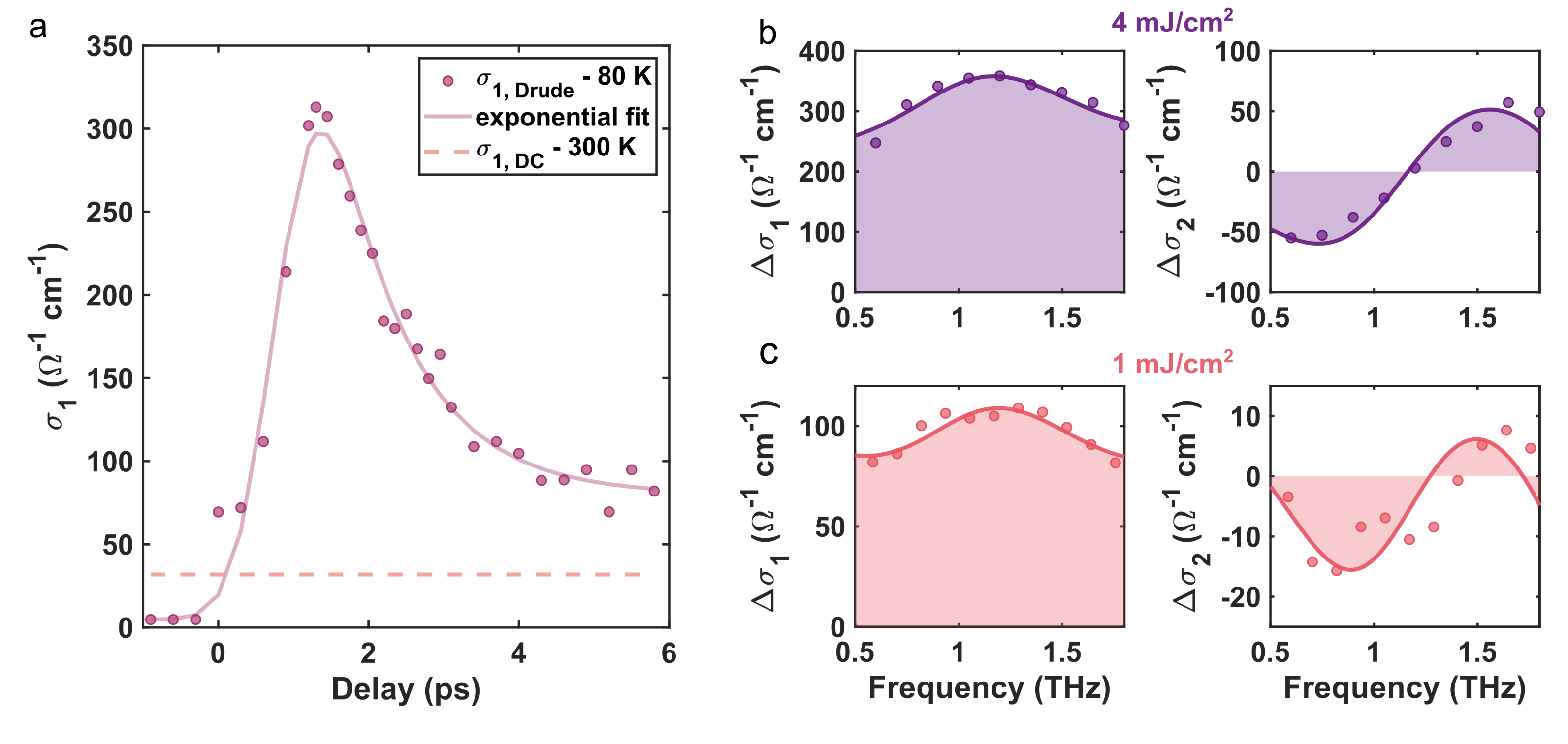}
\label{Fig4}
\end{figure}
\noindent
\textbf{Figure 4 $|$ Impact of free carriers on HEs in \327.} 
\textbf{a} Instantaneous DC conductivity of \327 extrapolated from the Drude component of fits to $\Delta\sigma_{1}(\omega)$ (Methods \cite{sm}) plotted versus time. Data were taken at a temperature of 80 K and a pump fluence of 4 mJ/cm$^{2}$. The purple curve is an exponential fit (Methods \cite{sm}). The reported equilibrium DC conductivity at 300 K (dashed line) \cite{cao_PhysRevB.66.214412} is shown for comparison. \textbf{b,c} $\Delta\sigma_{1}(\omega)$ (left column) and $\Delta\sigma_{2}(\omega)$ (right column) obtained with pump fluences of 4 mJ/cm$^{2}$ (c) and 1 mJ/cm$^{2}$ (d) at a temperature of 80 K and at $t$ = 1.9 ps. Solid lines are least-squared fits to a Drude-Lorentz model (Methods \cite{sm}). 

\newpage
%
%


\widetext

\begin{center}
\textbf{\large Supplementary Information}
\end{center}
\setcounter{equation}{0}
\setcounter{figure}{0}
\setcounter{table}{0}
\setcounter{page}{1}
\makeatletter
\renewcommand{\theequation}{S\arabic{equation}}
\renewcommand{\thefigure}{S\arabic{figure}}
\renewcommand{\bibnumfmt}[1]{[S#1]}
\renewcommand{\citenumfont}[1]{S#1} 

\section{Methods}
\subsection{Measurement of fixed THz electric field transients}
In Figure 3e, $\Delta E\left(t_{\mathrm{fixed}}, t \right)$ was measured at a fixed point in the time-domain THz pulse. This was done by anchoring the EOS gate pulse to a particular point $t_{\mathrm{EOS}} = t_{\mathrm{\mathrm{fixed}}}$ while $t$ was swept using the motorized delay stage in the pump path. While these fixed electric field pump-probe traces do not produce a frequency-resolved response, they provide frequency-integrated information about the dynamics when $t_{\mathrm{fixed}}$ is anchored to the peak of $E(t_{\mathrm{EOS}})$ \cite{poellmann_resonant_2015}. Importantly, these traces are over 100 times faster to acquire than the full mapping of $\Delta E(t_{\mathrm{EOS}}, t)$. To obtain the plot shown in Figure 3e, each pump-probe trace is then fitted to an exponential decay function convoluted with a Gaussian function (explained in more detail in section E.), and the pre-factor A is plotted as a function of temperature. Figure S5 shows the temperature-dependent pump-probe traces together with their exponential fits for both Sr$_{3}$Ir$_2$O$_{7}$ and Sr$_{2}$IrO$_{4}$.

\subsection{Calculation of the transient optical conductivity}
\subsubsection{Static optical conductivity}
In order to extract the photo-induced changes to the optical conductivity from the tr-TDTS measurement, an equilibrium optical response is needed. At low temperatures ($<$ 100 K), Sr$_{3}$Ir$_2$O$_{7}$ is transmissive enough to be measured using static TDTS in the transmission geometry, which allows for the experimetnal extraction of the equilibrium complex index of refraction $\tilde{n}\left(\omega\right)$, as shown in Figure S2a. In this case, ZnTe crystals were used for both generation of the THz pulse and EOS detection. The THz electric field was first measured by transmitting through the sample sitting over an aperture, with the light propagating along the (001) axis of the crystal. Then, the THz electric field transmitted through the bare aperture was measured, keeping all other parameters fixed. These two field transients were divided in the frequency domain to obtain the experimental complex transmission $\tilde{t}_{Exp.}\left(\omega\right)$. To extract $\tilde{n}\left(\omega\right)$, the difference between  $\tilde{t}_{Exp.}\left(\omega\right)$ and the expected theoretical response $\tilde{t}_{Th.}\left(\omega\right)=\ \frac{4\tilde{n}\left(\omega\right)}{\left[\tilde{n}\left(\omega\right)+1\right]^2}\times\left[\mathrm{exp}\frac{i\omega d}{c}\left(\tilde{n}\left(\omega\right)-1\right)\right]$  was minimized using a least-squares algorithm at each $\omega$ with $\tilde{n}\left(\omega\right)$ as the variable of interest \cite{duvillaret_1996}. Here $d$ is the sample thickness and $c$ is the speed of light in vacuum.

However, for both Sr$_3$Ir$_2$O$_7$ and Sr$_2$IrO$_4$, the sample becomes too metallic allow for THz transmission as the temperature is increased beyond 80 K. Thus, to obtain the static optical conductivity at all temperatures, we requested the real and imaginary optical conductivity data published in \cite{ahn.moon_srep16} for Sr$_3$Ir$_2$O$_7$ and \cite{seo_infrared_2017} for Sr$_2$IrO$_4$, which were obtained with Fourier transform infrared spectroscopy. For Sr$_2$IrO$_4$, the data was easily extrapolated to zero frequency after fitting with the RefFIT program \cite{kuzmenko_kramerskronig_2005}. For Sr$_3$Ir$_2$O$_7$, the increased conductivity above $T_N$ had to be accounted for. To do this, we digitized the DC transport data published in \cite{cao_PhysRevB.66.214412} to extract the real part of the conductivity at $\omega = 0$, setting the imaginary part to be zero. Since the conductivity at the lowest frequency measured in \cite{ahn.moon_srep16} is very close to the DC conductivity, we performed a linear interpolation to obtain the values in the intermediate frequency range (Figure S2b,c). As can be seen in Figure S2, at 80 K this extrapolation procedure produces a result that is similar to the response extracted from our static TDTS measurement.

\subsubsection{Transient optical conductivity}
The transient changes to the index of refraction $\Delta \tilde{n}(\omega)$, and consequently the optical conductivity, can be determined from $\frac{\Delta \tilde{E}(\omega)}{\tilde{E}(\omega)}$ since it is equal to $\frac{\Delta \tilde{r}(\omega)}{\tilde{r}(\omega)}$, which depends on $\Delta\tilde{n}(\omega)$ \cite{hunt_manipulating_2015}. We use the thin film approximation \cite{hunt_manipulating_2015} in which the pump-induced portion of the sample can be considered as a thin film on top of the un-excited bulk with a homogeneous index of refraction $\tilde{n}^{\prime}\left(\omega\right)=\tilde{n}\left(\omega\right)+\Delta n(\omega)$. This approximation is justified because the penetration depth mismatch exceeds an order of magnitude for both \214 and \327 at all temperatures \cite{ahn.moon_srep16, seo_infrared_2017}. 

An electrodynamic analysis of this situation using Maxwell’s equations yields an analytic solution for the transient changes to the optical conductivity \cite{hunt_manipulating_2015}: 

\begin{equation}
    \Delta\tilde{\sigma}(\omega)= \left(\frac{1}{377\times d}\right)\frac{\frac{\Delta \tilde{E}(\omega)}{\tilde{E}(\omega)}\left(\tilde{n}^{2}(\omega)-1\right)}{\frac{\Delta \tilde{E}(\omega)}{\tilde{E}(\omega)}\left[\cos(\theta_{0})-\sqrt{\tilde{n}^{2}(\omega)-\sin^{2}(\theta_{0})}\right]+2\cos(\theta_{0})}
\end{equation}
where $d$ is the penetration depth of the pump pulse and $\theta_{0}$ is the angle of incidence. 

\subsection{Drude-Lorentz Fitting}
Both the real and imaginary parts of the transient optical conductivity, $\Delta\sigma_{1}(\omega)$ and $\Delta\sigma_{2}(\omega)$, were simultaneously fit at each $t$ with the following Drude-Lorentz model: 
\begin{align}\label{dlfitfunction}
D\left[\frac{1}{\gamma_{\mathrm{Drude}}-i\omega}\right] 
&+ L_{\mathrm{HE}}\left[\frac{\omega}{i\left({\omega_{\mathrm{HE}}}^2\ -\omega^2\ \right)+\omega \gamma_{\mathrm{HE}}}\right] \nonumber \\
&+ L_{\mathrm{phon.}}\left[\frac{\omega}{i\left({\omega_{\mathrm{phon.}}}^2\ -\omega^2\ \right)+\omega \gamma_{\mathrm{phon.}}}\right] \nonumber \\
&+ L_{\mathrm{Bgd.}}\left[\frac{\omega}{i\left({\omega_{\mathrm{Bgd.}}}^2\ -\omega^2\ \right)+\omega \gamma_{\mathrm{Bgd.}}}\right]
\end{align}\\
The first term is a Drude term. The first of the three Lorentzian terms represents the HE mode. The second Lorentzian term captures the lowest frequency phonon mode in Sr$_{3}$Ir$_2$O$_{7}$, which has a significant spectral overlap with the exciton mode and has to be explicitly included in the fitting process to obtain a high-quality fit. The final Lorentzian term describes a weak background, likely caused by pump-induced changes of higher energy phonon transitions. The fitting constants $D$, $L_{\textrm{HE}}$, $L_{\textrm{phon.}}$, and $L_{\textrm{Bgd.}}$ are the strengths of the Drude, HE, lowest energy phonon, and background terms, respectively. The fitting constants $\gamma_{\textrm{Drude}}$, $\gamma_{\textrm{HE}}$, $\gamma_{\textrm{phon.}}$, and $\gamma_{\textrm{Bgd.}}$ are the widths of the Drude, HE, phonon, and background terms, respectively. The fitting constants $\omega_{\textrm{HE}}$, $\omega_{\textrm{phon.}}$, and $\omega_{\textrm{Bgd.}}$ are the central frequencies of the HE, phonon, and background terms, respectively. 
Since the Drude-to-HE crossover is observed in Sr$_2$IrO$_4$ at all temperatures measured, we used the same fitting process for all temperatures. Since there is no phonon mode very close to the exciton mode in frequency, only the Drude, HE, and background terms are included in the fits. For $t \leq 450$ fs, the value of $\gamma_{\textrm{Drude}}$ was left as a free parameter, while for later $t$ it was fixed to its average fitted value in order to constrain the number of free parameters and improve the quality of the fit. No HE component was included for $t \leq 450$ fs due to the absence of the HE Lorentzian at these delays.

In the case of Sr$_{3}$Ir$_2$O$_{7}$, the fitting process is slightly different for $T$ $<$ \TN, where a Drude-to-HE crossover is observed as a function of $t$, and for $T > $ \TN\ where only a positive metallic response is observed. In the former case, for $t \leq 300$ fs, only the Drude term was included in the fit and the value of $\gamma_{\textrm{Drude}}$ was left as a free parameter. For later $t$, we included all terms in Equation \ref{dlfitfunction} and $\gamma_{\textrm{Drude}}$ was fixed to its average fitted value in order to constrain the number of free parameters and improve the quality of the fit. In the latter case, we found that the Drude model did not fit the data. This behavior is possibly to the correlated nature of Sr$_{3}$Ir$_2$O$_{7}$, which introduces an incoherent character to the transport signatures \cite{foulquier_evolution_2023}. Nonetheless, both the real and imaginary parts of $\Delta\sigma$ are positive, which is similar to a Drude response and indicates metallic behavior, and there is clearly an absence of the Lorentzian mode at the HE frequency seen in the case of $T$ $<$ \TN. For these reasons, we did not plot a fit to a Drude-Lorentz model with the data acquired at or above $T_{N}$ in \327. 

These procedures result in a high quality of fit for all spectra as shown in Figure 2b, Figures 3a-c, Figures 4b-c, and Figure S6. The spectral weights in Figure 2c are calculated by integrating over each term in the fit individually, following the partial oscillator strength sum rule \cite{kaindl_transient_2009}. In Figure S3, a representative data set is shown along with each of the individual components of the fit.

Finally, we discuss the origin of the phonon and background terms. 
The phonon response in our spectrum is due to an energy transfer from the electronic subsystem to the lattice subsystem, as is typical for systems excited by photons with energies that lie above the band gap. The phonons clearly do not redshift upon optical excitation because their frequencies in the photo-induced response are identical to those in the equilibrium response, as shown in Figure S1. Moreover, redshifting results in a distinct first derivative-like, negative-positive structure in the transient reflectivity spectrum (such as the $\Delta R/R$ spectrum in La$_2$CuO$_4$ upon photodoping \cite{baldini_2020_electron}), which we do not observe in our $|\Delta E/E|$ spectrum in the phonon frequency range. 

The background term simply represents the combined response from all other higher-energy phonons and charge excitations. These Lorentzian modes have long tails towards low frequencies in the imaginary part, which we need to account for explicitly in the fitting process. Consistent with the typical protocol in optical conductivity analysis, we choose to use a single Lorentzian mode to approximate the higher-frequency response to limit the number of fitting parameters and maintain a Kramers–Kronig consistent fitting function.

\subsection{Exponential Fitting}
To fit the data in Figure 2c and Figure S5, we used an exponential function convolved with a Gaussian function, which explicitly reduces to the form: 

\begin{equation}
    \left[1+\ \mathrm{erf}\left(\frac{t-t_0}{t_r}\right)\right]\times\left[A\ \mathrm{exp} \left(-\frac{t-t_0}{\tau}\right)+b\right]
\end{equation}\\
\noindent
where $t$ is the pump-probe time delay, and the fitting parameters $t_{0}$, $t_{r}$, $b$, $A$ and $\tau$ are time-zero, the rise time, the pump-induced offset, and the strength and decay constant of the exponential respectively. When the exponential decay is slower than the rise time (i.e. when $\tau > t_r$) $t_0$ is approximately located at the halfway of the rising edge. This is the case for Sr$_{2}$IrO$_{4}$ at all temperatures and Sr$_{3}$Ir$_2$O$_{7}$ at $T$ $<$ \TN. For Sr$_{3}$Ir$_2$O$_{7}$ at $T$ $>$ \TN, Since $\tau \approx t_r$, we constrained $t_0$ to before the maximum of the trace to obtain a stable fit. 

\subsection{Excitation density calculations}
The excitation density is given by the number of pump photons absorbed in each pulse, divided by the total number of Ir-O layers in the volume illuminated by the pump beam. This is calculated with the formula: 

\begin{equation}
    \rho = \frac{F(1-R)}{\hbar\omega}\frac{d}{c_{0}/N_{l}}
\end{equation}

\noindent where $R$ is the reflectivity at the pump wavelength, $\hbar\omega$ is the pump photon energy, $c_{0}$ is the lattice constant along the c-axis (perpendicular to the Ir-O planes), $d$ is thickness given by the penetration depth at the pump wavelength, and $N_{l}$ is the number of Ir-O layers in a unit cell.

The excitation fluence $F$ is calculated by: 
\begin{equation}
    F = \frac{P}{f_{rep} A}
\end{equation}
where $P$ is the average power of the pump beam, $f_{rep}$ is the repetition rate, and $A$ is the area of the pump spot calculated with the FWHM spot size. 

\subsection{Estimation of the static laser heating effect}
We calculated the temperature increase due to steady-state laser heating at the maximum fluences used in the experiments (2 mJ/cm$^2$ for \214 and 4 mJ/cm$^2$ for \327) using the formula $\Delta T = \frac{P d}{A \kappa}$, where the thermal conductivity $\kappa$ is digitized from \cite{pallecchi_2016_thermoelectric}, P is the laser power absorbed by the sample, d is the thickness of the samples, and A is the spot size of the laser beam defined at the FWHM. This calculation gives us a temperature increase due to static laser heating effect of less than 2 K for both compounds at all temperatures, which is less than the temperature steps used in the measurements shown in Figure 3. Since the temperature increase is negligible, we did not apply any correction to the temperature.

\newpage
\section{Extended Data Figures}
\subsection*{Figure S1: Broadband equilibrium optical response of \327}
\begin{figure}[h]
\includegraphics[width=90mm]{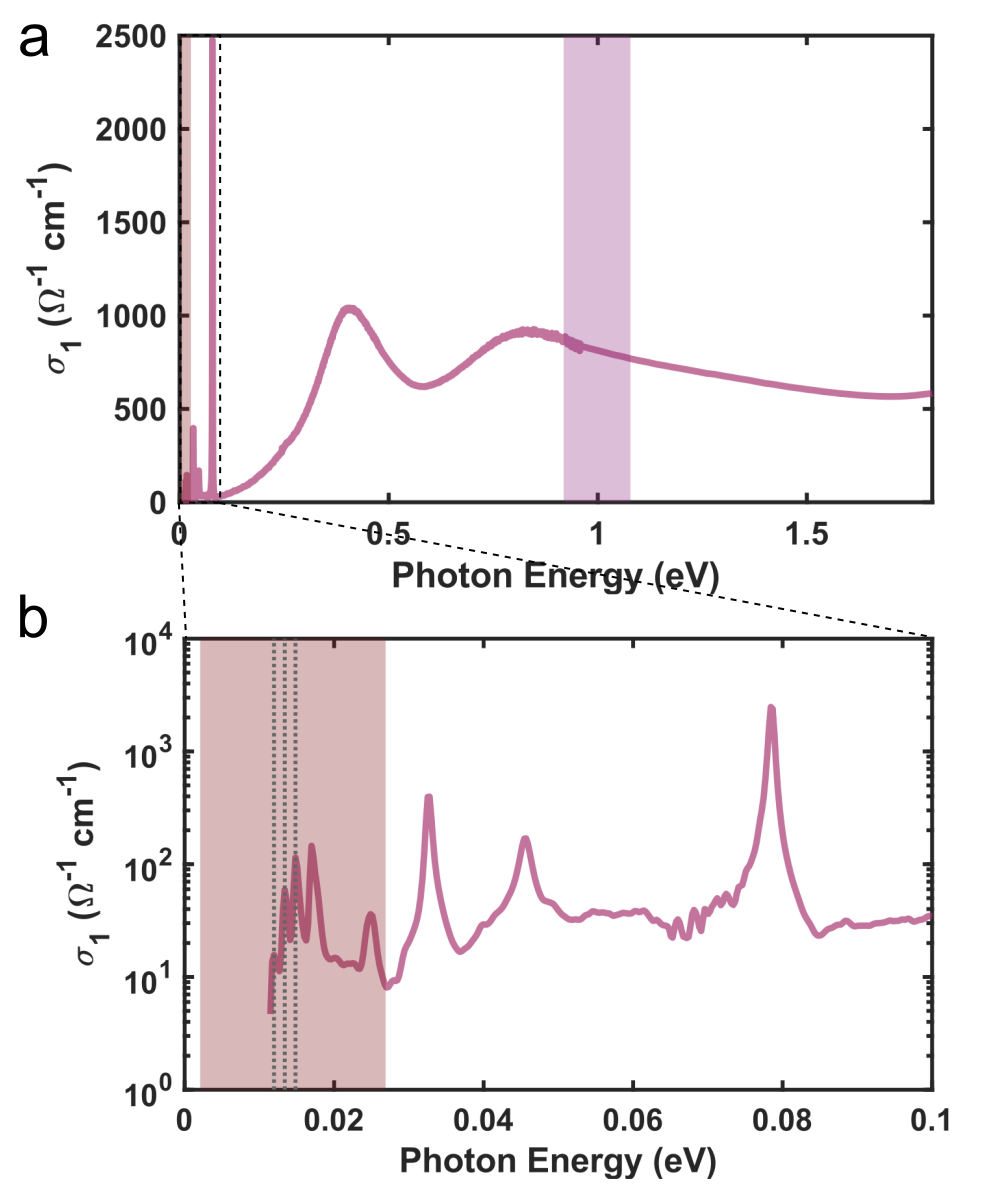}
\label{ExFig1}
\end{figure}
\noindent
\textbf{a} Equilibrium optical conductivity of \327 at 100 K \cite{ahn.moon_srep16}. The maroon and purple rectangles indicate the frequencies of the THz probe and NIR pump respectively. The sharp peaks below 0.1 eV are phonons while the broad peaks at 0.4 eV and 0.8 eV are the $\alpha$ and $\beta$ electronic transitions, respectively. \textbf{b} Same as panel a but focused on the low-frequency range that encompasses the phonon peaks \cite{ahn.moon_srep16}. The dotted lines indicate the frequencies of the phonons observed in our time-resolved THz spectrum (Fig. 2a), which agree well with the phonon modes observed in the equilibrium optical conductivity spectrum. Note that the bandwidth of the spectrum in \cite{ahn.moon_srep16} did not include the lowest 2.26 THz phonon that we observed. 
\newpage

\subsection*{Figure S2: Equilibrium THz conductivity of \327}
\begin{figure}[h]
\includegraphics[width=140mm]{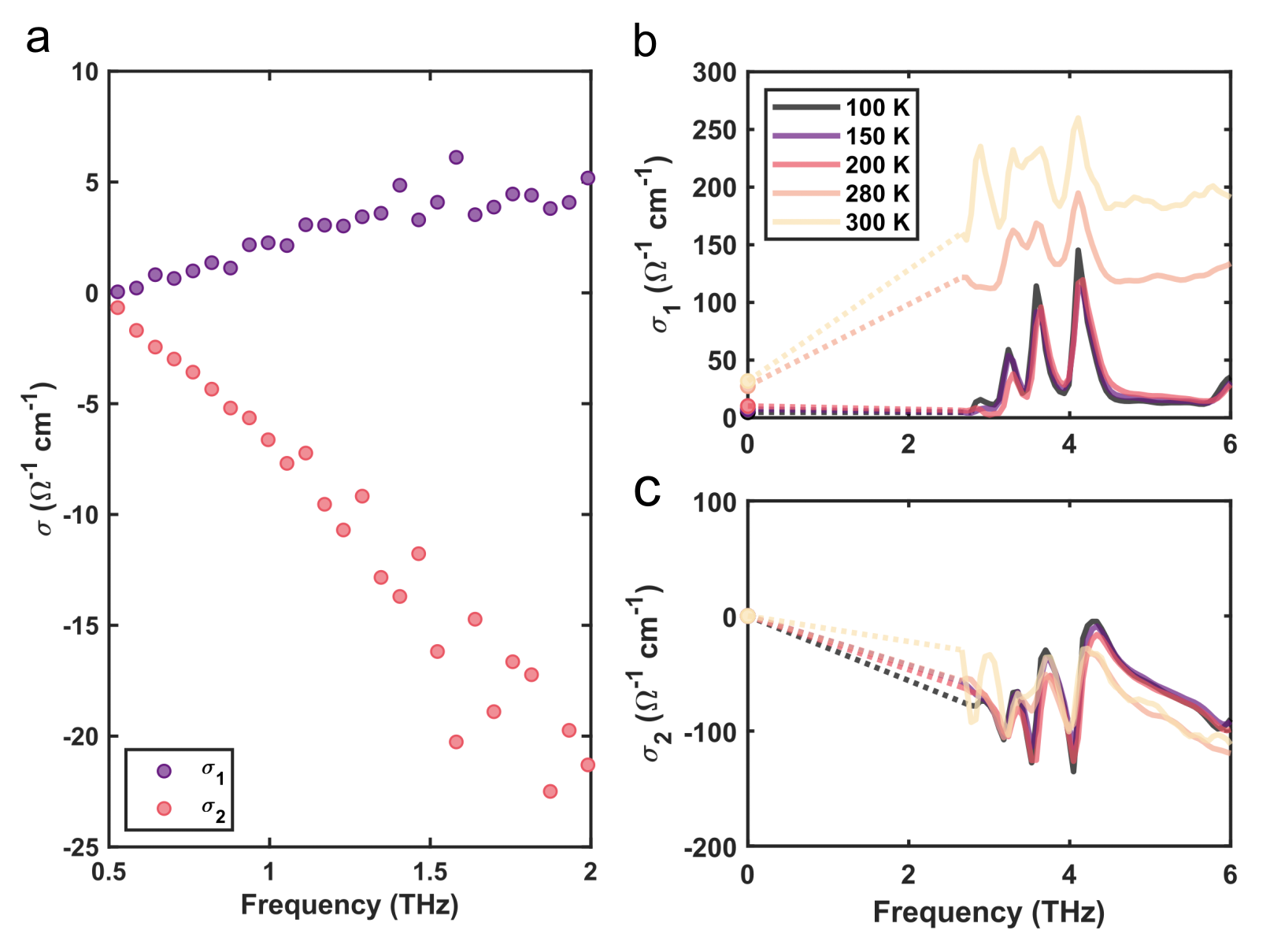}
\label{ExFig2}
\end{figure}
\noindent
\textbf{a} Real $\sigma_{1}(\omega)$ and imaginary $\sigma_{2}(\omega)$ parts of the optical conductivity of \327 in equilibrium at 80 K obtained from TDTS measurements performed in transmission geometry (Methods). Both $\sigma_{1}(\omega)$ and $\sigma_{2}(\omega)$ are featureless and no Lorentzian mode is seen. \textbf{b,c}  $\sigma_{1}(\omega)$  (b) and  $\sigma_{2}(\omega)$ (c) obtained from Fourier transform infrared spectroscopy (FTIR) measurements \cite{ahn.moon_srep16}. Solid circles are the DC conductivity \cite{cao_PhysRevB.66.214412}. Dashed lines are the linear interpolation between the DC conductivity and the FTIR results. The spectrum shown in panel a is consistent with the low-temperature optical conductivity from the literature shown in panels b and c.  
\newpage

\subsection*{Figure S3: Drude-Lorentz fitting of a typical tr-TDTS spectrum of \327}
\begin{figure}[h]
\includegraphics[width=140mm]{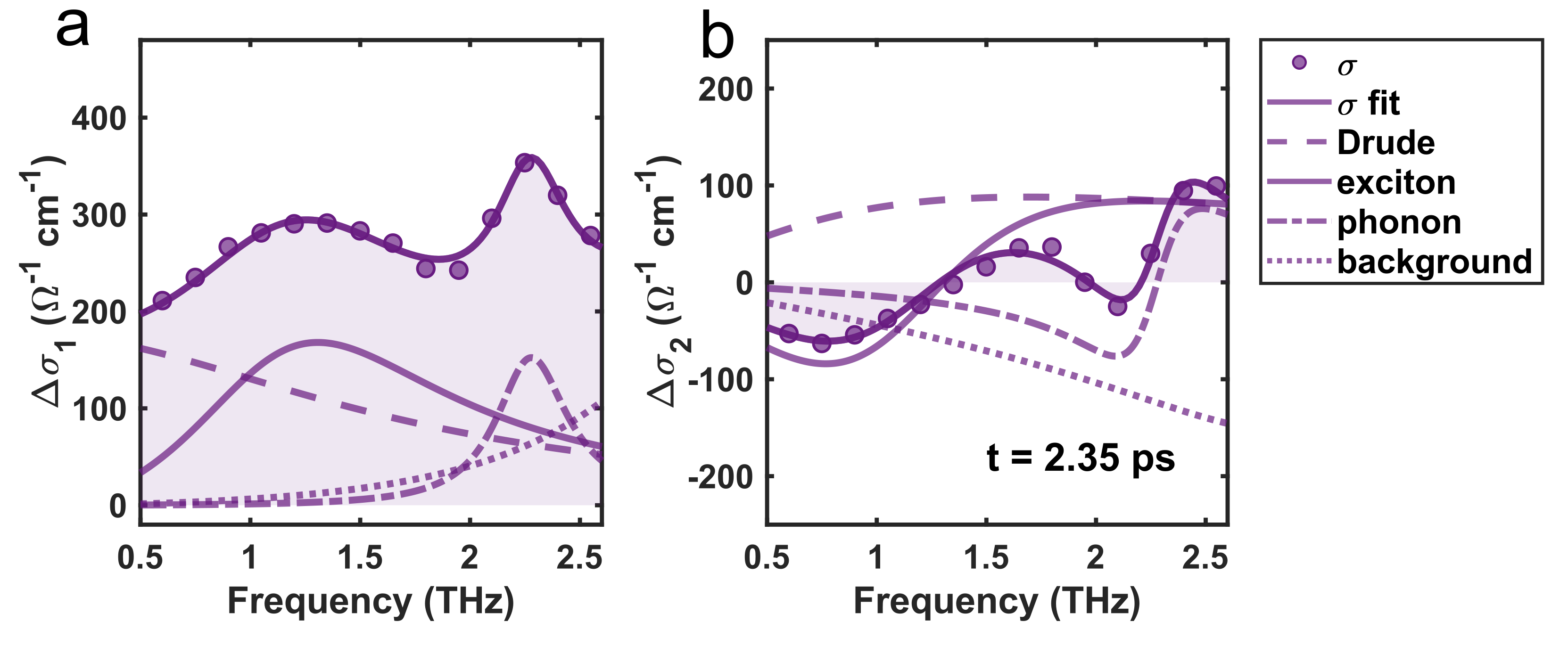}
\label{ExFig3}
\end{figure}
\noindent
\textbf{a,b} Individual components of the Drude-Lorentz fitting of $\Delta\sigma_{1}(\omega)$ (a) and $\Delta\sigma_{2}(\omega)$ (b) plotted with the data at $t = 2.35$ ps for \327. The frequency range for the fit is chosen to include both the HE and the lowest frequency phonon to obtain a high-quality fit. The pump photon energy was fixed at 1.0 eV and set to a fluence of 4 mJ/cm$^{2}$. The temperature of the sample was 80 K. 
\newpage

\subsection*{Figure S4: Dynamics of the Drude-like response in \327 at $T$ $>$ \TN}
\begin{figure}[h]
\includegraphics[width=\textwidth]{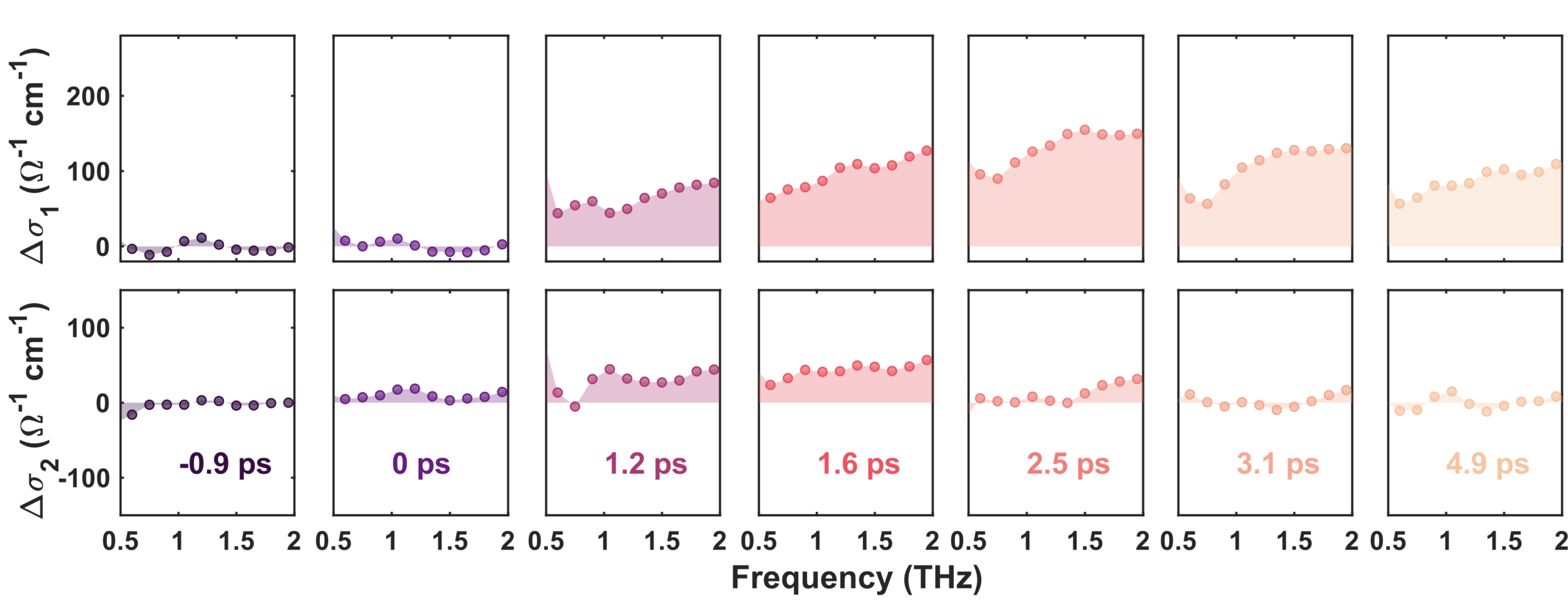}
\label{ExFig4}
\end{figure}
\noindent
$\Delta\sigma_{1}(\omega)$ (upper row) and $\Delta\sigma_{2}(\omega)$ (lower row) at various $t$ extracted from $\Delta E/E$  (Methods), displayed as dots and shadings. The system shows a Drude-like response at all $t$, and no Lorentzian-like feature is observed at any $t$. The data were acquired at a temperature of 300 K with a pump fluence of 4 mJ/cm$^{2}$.

\newpage
\subsection*{Figure S5: Temperature dependence of the frequency-integrated photo-induced change to the THz response.}
\begin{figure}[h]
\includegraphics[width=140mm]{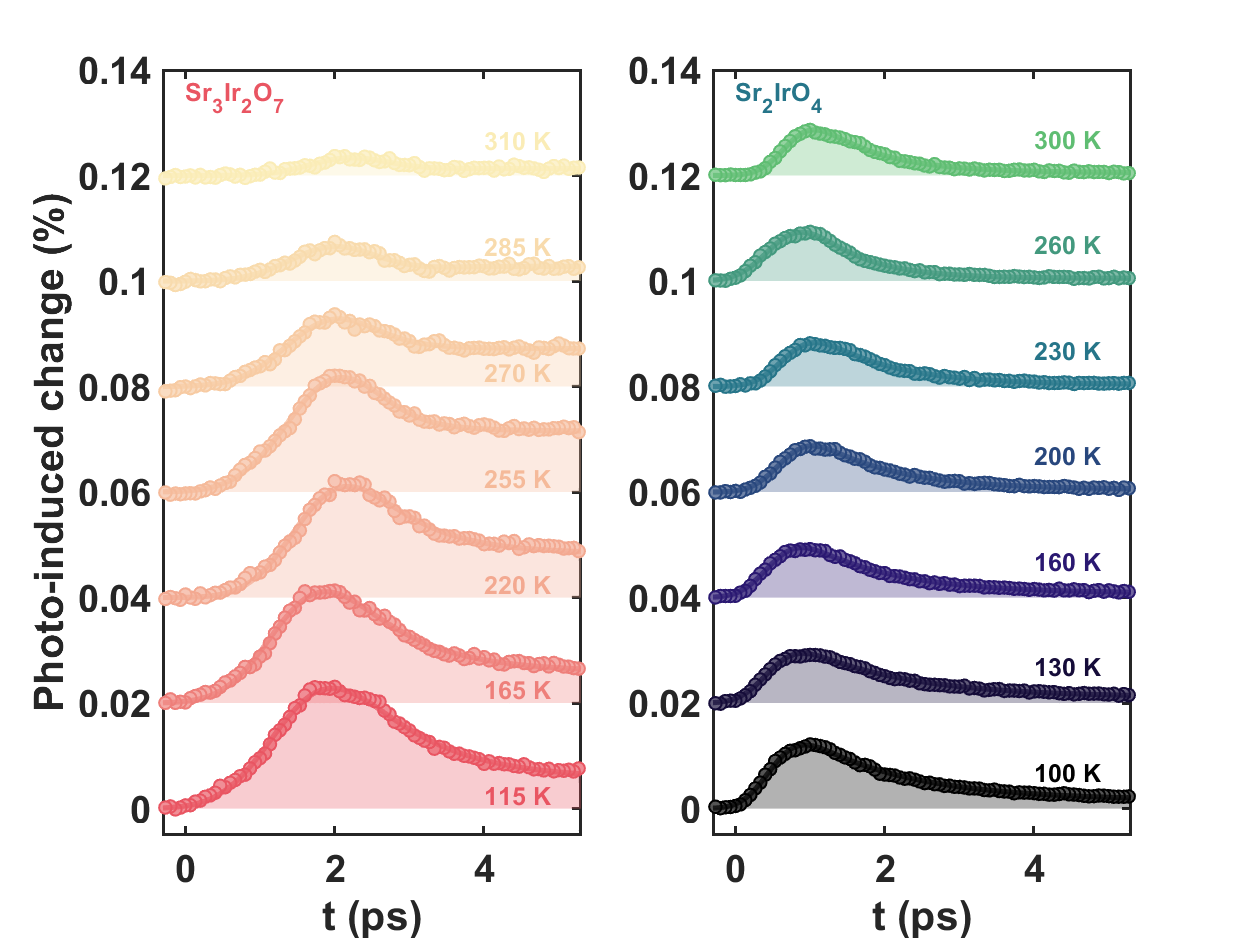}
\label{ExFig5}
\end{figure}
\noindent
\textbf{a-b,} Frequency-integrated response of $\Delta E/E$ (Methods) measured as a function of $t$ at a series of temperatures for (left) Sr$_{3}$Ir$_{2}$O$_{7}$ (pump photon energy = 1.0 eV, fluence = 4 mJ/cm$^2$) and (right) Sr$_{2}$IrO$_{4}$ (pump photon energy = 0.6 eV, fluence = 2 mJ/cm$^2$). The solid lines and shaded areas show the fits to an exponential decay function convolved with a Gaussian (Methods). 
\newpage

\subsection*{Figure S6: Fluence dependence of the transient THz spectrum of \214}
\begin{figure}[h]
\includegraphics[width=100mm]{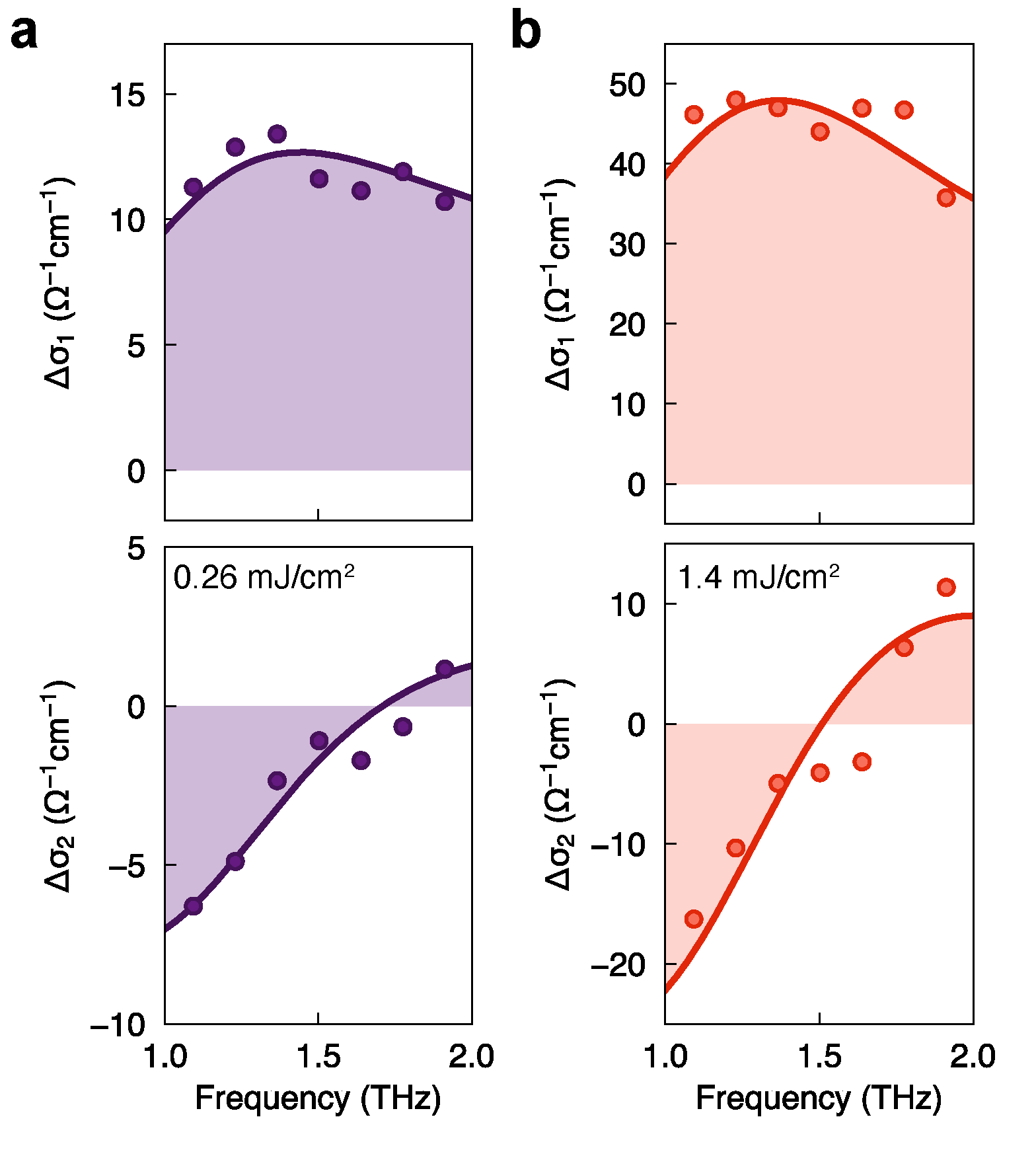}
\label{ExFig6}
\end{figure}
\noindent
\textbf{a,b} Real (top) and imaginary (bottom) parts of the photo-induced changes to the optical conductivity obtained with pump fluences of 0.26 mJ/cm$^2$ (a) and 1.4 mJ/cm$^2$ (b). These data were taken at $t$ = 2.55 ps and a temperature of 80 K. The pump photon energy was tuned to 0.6 eV. These data are reproduced from \cite{Mehio_Hubbard_2023}.
\newpage

\subsection*{Figure S7: Temperature dependence of the HE and phonon response in \327}
\begin{figure}[h]
\includegraphics[width=\textwidth]{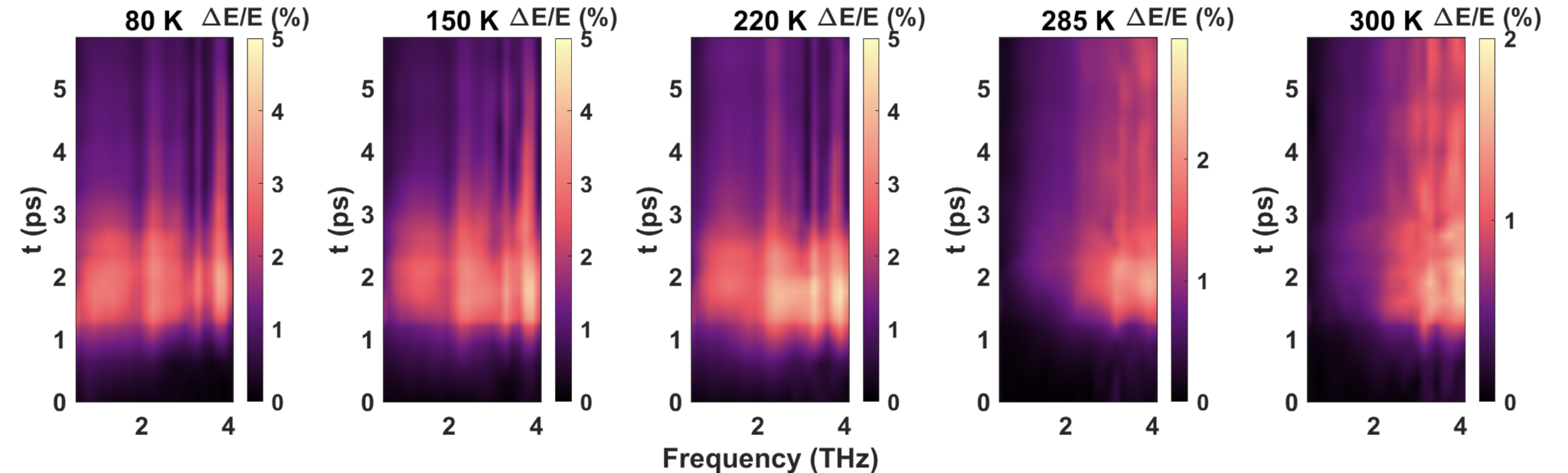}
\label{ExFig7}
\end{figure}
\noindent
Pump-induced change of the broadband THz spectra as a function of pump-probe delay $t$ showing both the HE response below 2 THz and the phonon responses above 2 THz. The colors represent the magnitude of $\Delta E/E$. The data were acquired with a pump fluence of 4 mJ/cm$^{2}$ and at a range of temperatures from 80 K to 300 K, as indicated in the figure. Note that the color scales are different for each temperature such that the phonons appear as roughly the same color in each. As the temperature increases above \TN\ = 285 K, the phonon peaks persist (although they become weaker and broader) whereas the exciton peak disappears. 
\newpage

\subsection*{Figure S8: Temperature dependence of the HE central frequency}
\begin{figure}[h]
\includegraphics[width=\textwidth]{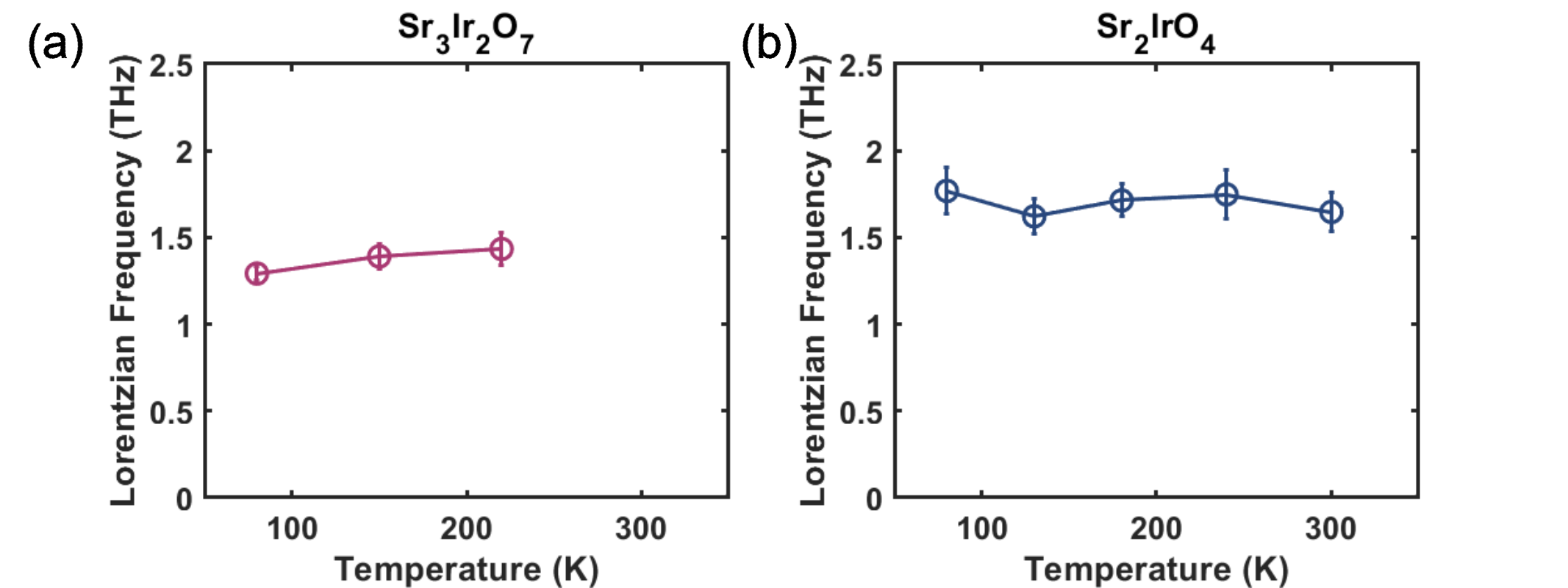}
\label{ExFig8}
\end{figure}
\noindent
The central frequency of the HE mode as a function of temperature in (a) \327 and (b) \214, at representative delays of 2.2 ps and 1.8 ps respectively, obtained from fitting to a Drude-Lorentz model (Section I.D.). The error bars are the standard deviation from the least-squares fitting algorithm.
\newpage

%

\end{document}